\documentclass[12pt]{iopart}

\usepackage{graphicx}

\begin{document}

\title{Influence of Preparation Process on Microstructure, Critical Current Density and T$_{c}$ of MgB$_{2}$/Fe/Cu
Wires}

{Y.~F.~Wu$^1$\footnote[1]{To whom correspondence should be addressed
(wyf7777@tom.com)}, G.~Yan$^1$, J.~S.~Li$^2$, Y.~Feng$^1$,
S.~K.~Chen$^1$, H.~P.~Tang$^1$, H.~L.~Xu$^3$, C.~S.~Li$^1$, P.~X.~
Zhang$^1$, Y.~F.~Lu$^1$}

\address{$^1$ Northwest Institute for Nonferrous Metal Research, P. O. Box 51, Xian, Shaanxi 710016, P. R. China}
\address{$^2$ Northwestern Polytechnical University, Xi'an 710012, P.R.China}
\address{$^3$ School of Material Science and Engineering, Zhengzhou University, Zhengzhou, Henan 450002, People¡¯s Republic of
China}

\begin{abstract}
The powder-in-tube MgB$_{2}$ wires were prepared by high energy
milling of Mg and B powder. The powder was not mechanically alloyed
for 2h short milling time. However, the MgB$_{2}$ grains in wires
are very small (20~100nm) and resembled to the dimple after
post-heat treatment. The clear evidence for transcrystlline fracture
is observed. It indicated that the grain connection was greatly
improved and the fluxing pinning was significantly enhanced. Another
point to view is no intermediate annealing was adopted during the
whole rolling process. The influence of the post-heat treatment on
the transport current density was studied. Despite the lower T$_{c}$
of about 35K, the transport current density reaches to
3$\times\,$10$^{4}$A/cm$^{2}$ at 15K and 3.5T for
$700\,^{\circ}\mathrm{c}$ sintered wires.

\end{abstract}

%Uncomment for PACS numbers title message
%\pacs{00.00, 20.00, 42.10}
% Keywords required only for MST, PB, PMB, PM, JOA, JOB?
%\vspace{2pc}
%\noindent{\it Keywords}: Article preparation, IOP journals
% Uncomment for Submitted to journal title message
%\submitto{\JPA}
% Comment out if separate title page not required

% PACS codes here, in the form: \PACS code \sep code

%74. Superconductivity (for superconducting devices, see 85.25.-j)

%74.25.Bt Thermodynamic properties

%74.25.Fy Transport properties (electric and thermal conductivity, thermoelectric effects, etc.)

%74.25.Ha Magnetic properties

%74.25.Ld Mechanical and acoustical properties, elasticity, and ultrasonic attenuation

%74.25.Nf Response to electromagnetic fields (nuclear magnetic resonance, surface impedance, etc.)

%74.25.Op Mixed states, critical fields, and surface sheaths

%74.25.Qt Vortex lattices, flux pinning, flux creep

%74.25.Sv Critical currents

%74.50.+r Tunneling phenomena; point contacts, weak links, Josephson effects (for SQUIDs, see 85.25.Dq; for Josephson devices, see 85.25.Cp; for Josephson junction arrays, see 74.81.Fa)

%74.62.-c Transition temperature variations

%74.62.Bf Effects of material synthesis, crystal structure, and chemical composition

%74.62.Dh Effects of crystal defects, doping and substitution

%74.62.Fj Pressure effects

%74.70.-b Superconducting materials (for cuprates, see 74.72.-h)

%74.70.Ad Metals; alloys and binary compounds (including A15, MgB2,etc.)

\maketitle

\section{Introduction}
Preparation of MgB$_{2}$ wires depends very critically on the
precursor powder used for the powder-in-tube (PIT) techniques. Small
grain sizes are favored preconditions for achieving high quality
wires. Mechanical alloying (MA) technique for MgB$_{2}$ powder
preparation is expected for obtaining enhanced magnetic flux pinning
by microstructure refinement. However, it costs as long as
20~100h[1,2,3-8] for in situ MA precursor powder preparation. The
use of short-time unalloyed high energy milling of Mg and B powder
as precursor material represents an efficient combination between
conventional powder preparation and mechanical alloying techniques.
Fe sheath was considered as one of suitable materials however hard
it is. It is also gradually recognized that repeated annealing
during rolling process would cause the inevitable diffusion of
oxygen into the filament of MgB$_{2}$ wire and decrease J$_{c}$
greatly. Therefore, how to reduce intermediate annealing steps
during material preparation become one of the biggest challenges as
literatures[8-10] referred to.

\section{Experimental details}
Mg(99.8\%) and amorphous B (95\%) powder with 5\% Mg surplus were
filled under purified Ar-atmosphere into an agate milling container
and milling media. The milling was performed on a SPEX 8000M mill
for 2h using a ball-to-powder mass ratio of 3. Monofilamentary wires
were prepared by conventional PIT method. The powder was packed into
coaxial Cu/Fe tubes (outer diameter 14 mm and inner diameter 7 mm)
forming the billets. Next, billets were groove-rolled and drawn to a
wire diameter of 1.2mm without any intermediate annealing. The post
heat treatment was performed at different temperatures for 1 hour
under ultra-high purity Ar-atmosphere. The phase content of
MgB$_{2}$ was analysed by x-ray diffraction scans performed on a
Philips APD1700 diffractometer with Cu K¦Á radiation. The surface
morphology and microstructures of the samples were characterized by
JSM-6700F scanning electron microscope. The superconducting
transition temperature, T$_{c}$, was obtained by
resistance-temperature method. The critical currents were evaluated
from V-I curves taking a 1¦ÌV cm$^{-1}$ criterion.

\section{Results and discussion}
As we know, the ductile deformation behavior at low temperatures is
favored for acquiring high J$_{c}$ performance MgB$_{2}$ wires. In
our study, no intermediate annealing was adopted during the whole
rolling process. The total working modulus of Fe sheath exceeds 99
percent. A typical cross section for wire diameter of 1.2 mm is
shown in figure 1. We concluded from our experience that the
reduction in pass and rolling rate during rolling process should be
controlled to a lower lever.

The microstructures of samples sintered at different temperatures
are shown in Fig.2. The Scanning electron microscope images mainly
show dimple-like grains for $700~800\,^{\circ}\mathrm{c}$ sintered
samples. The grain sizes are about 20 ~ 100nm for
$700\,^{\circ}\mathrm{c}$ and $750\,^{\circ}\mathrm{c}$ sintered
samples. The clear evidence for the transcrystlline fracture is
observed. It indicated that the grain connectivity was enhanced
greatly. The grains are grown up to 100~250nm for
$800\,^{\circ}\mathrm{c}$ sintered sample. While for
$650\,^{\circ}\mathrm{c}$ reacted sample, large plate-like column
crystals were observed. The impurity phases are evidently observed
for sample sintered at $800\,^{\circ}\mathrm{c}$. Fig.3 shows the
wrap-around distribution of the second phases of the
$800\,^{\circ}\mathrm{c}$ sintered sample. EDX analysis indicates
that the bright one is O-rich zone and the grey one is Mg-rich
zone,as seen in fig.4.

Fig.5 indicates the superconducting transition T$_{c}$ by resistance
measurement for the samples sintered at different temperatures. As
we can see, all samples have sharp transitions. T$_{c}$ of the high
energy milling samples ranged from 34.5 to 35.5 K, which is much
lower than 39K. Most likely, this is not due to a deviation from the
ideal stoichiometry of the superconducting compound[2]. From x-ray
analysis, interaction between the constituents of precursor powder
and the sheath material can also be excluded. It seems that the
suppression of T$_{c}$ is caused by oxygen contamination in grain
boundary induced by high energy milling as well as sintering
process.

Fig.6 shows the \textit{J$_{c}$-{B}} curve of the MgB$_{2}$ wires at
different temperatures. The critical currents were evaluated from
V-I curves taking a 1¦ÌV cm$^{-1}$ criterion. As we can see, the
critical current densities are influenced greatly by heat treatment.
The higher or lower heat treatment induced second phases and
inferior microstructure, which led to bad grain connectivity and
contribute to dramatic decrease of \textit{J$_{c}$}. In the lower
field, the MgB$_{2}$ wires sintered at $700\,^{\circ}\mathrm{c}$
show the higher critical current density. In the higher field, the
MgB$_{2}$ wires sintered at $750\,^{\circ}\mathrm{c}$ show the
higher critical current density. The critical current density
reaches to 3$\times\,$10$^{4}$A/cm$^{2}$ at 15K and 3.5T for the
wire sintered at $700\,^{\circ}\mathrm{c}$.

In Summary, we succeeded in preparing high critical current density
MgB$_{2}$ wire using unalloyed high energy milling precursor powder.
The powder preparation process was greatly shortened. It
demonstrated that it is an effective approach to get fine
crystalline MgB$_{2}$ with good grain connectivity and high
\textit{J$_{c}$} performance. The critical current density reaches
to 3$\times\,$10$^{4}$A/cm$^{2}$ at 15K and 3.5T for the wire
sintered at $700\,^{\circ}\mathrm{c}$.

\section{Acknowledge}
This work was partially supported by National Natural Science
Foundation project (contract No. 50472099) and National Basic
Research Program of China (contract No. 2006CB601004).

\newpage
\textbf{Figure captions}

Fig. 1  A typical cross section for wire diameter of 1.2 mm.

Fig.2 The Scanning electron microscope images of samples sintered at
different temperatures: (a)$650\,^{\circ}\mathrm{c}$;
(b)$700\,^{\circ}\mathrm{c}$; (c)$750\,^{\circ}\mathrm{c}$ and
(d)$800\,^{\circ}\mathrm{c}$.

Fig.3  The wrap-around distribution of the second phases for
$800\,^{\circ}\mathrm{c}$ sintered MgB$_{2}$ wires.

Fig.4 The EDX analysis of the second phases for
$800\,^{\circ}\mathrm{c}$ sintered MgB$_{2}$ wires. It shows: (a)
the bright one is O-rich zone; (b) the grey one is Mg-rich zone.

Fig.5  Superconducting transition T$_{c}$ by resistance measurement
for the samples sintered at different temperatures.

Fig.6  Transport \textit{J$_{c}$-B} curve of the MgB$_{2}$ wires at
different temperatures.

\newpage
Fig.1
%\section{A typical cross section for wire diameter of 1.2mm}
%%%%%%%%%%%%%%%%%%%%%%%%  FIGURE 1  A typical cross section for wire diameter of 1.2mm  %%%%%%%%%%%%%%%%%%%%%%%%%
\begin{figure}[hp]
\centering
\includegraphics[width=200pt]{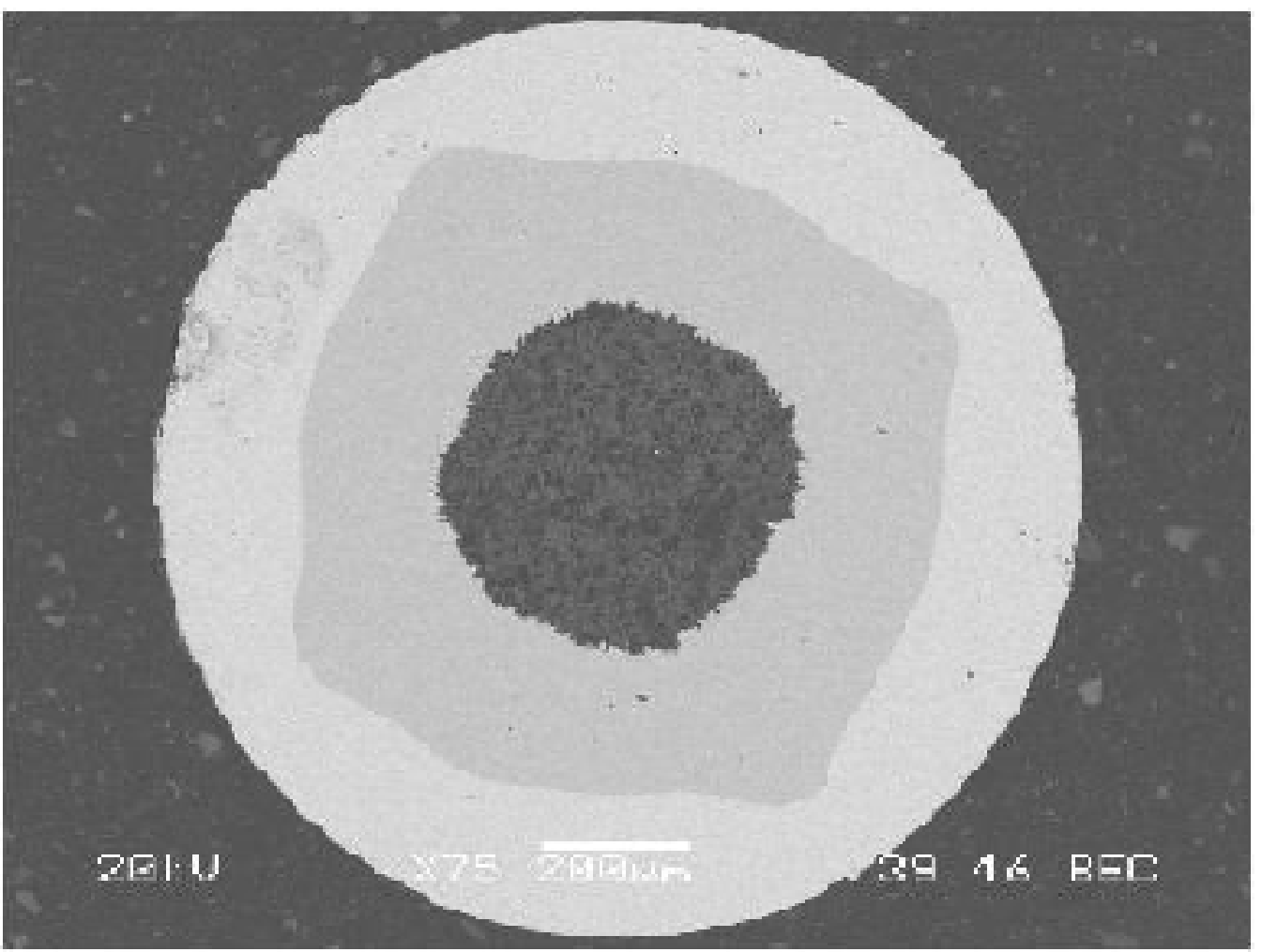}
\end{figure}
%%%%%%%%%%%%%%%%%%%%%%%%%%%%%%%%%%%%%%%%%%%%%%%%%%%%%%%%%%%%

\newpage
Fig.2
%\section{SEM for diff temperatures}
%%%%%%%%%%%%%%%%%%%%%%%%  FIGURE 2 SEM for diff temperatures %%%%%%%%%%%%%%%%%%%%%%%%%
\begin{figure}[hp]
\centering
\includegraphics[width=200pt]{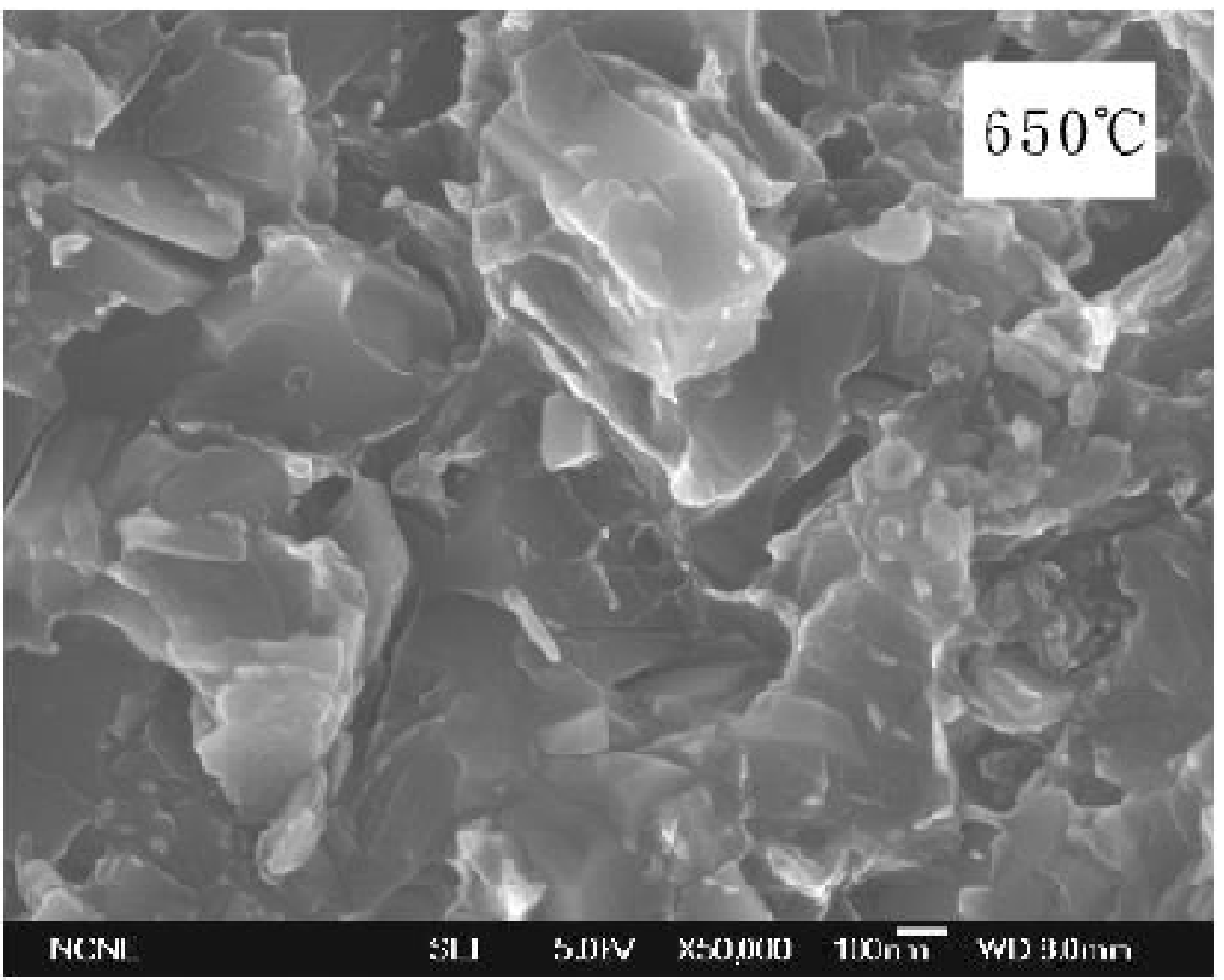}%
\includegraphics[width=200pt]{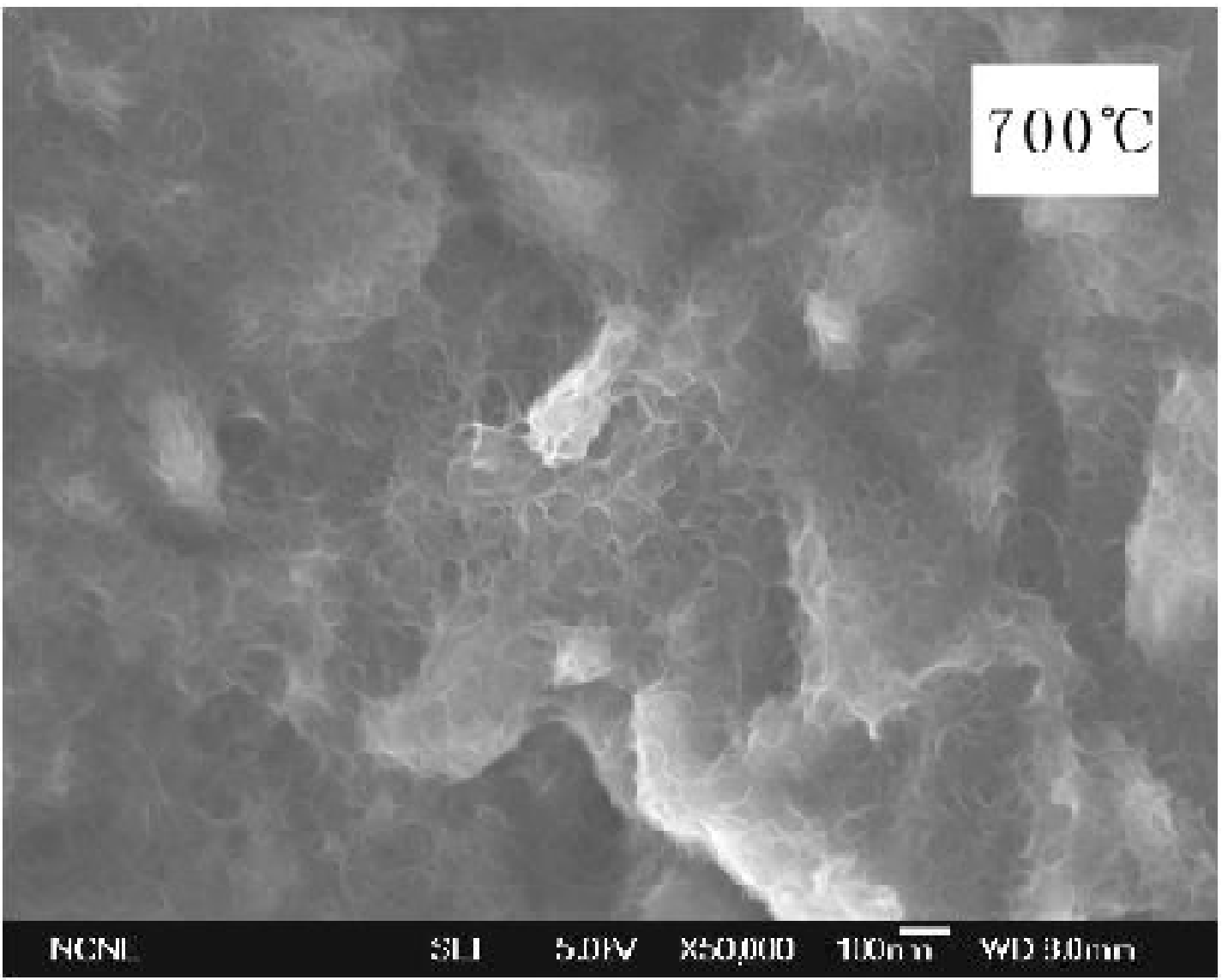}
\includegraphics[width=200pt]{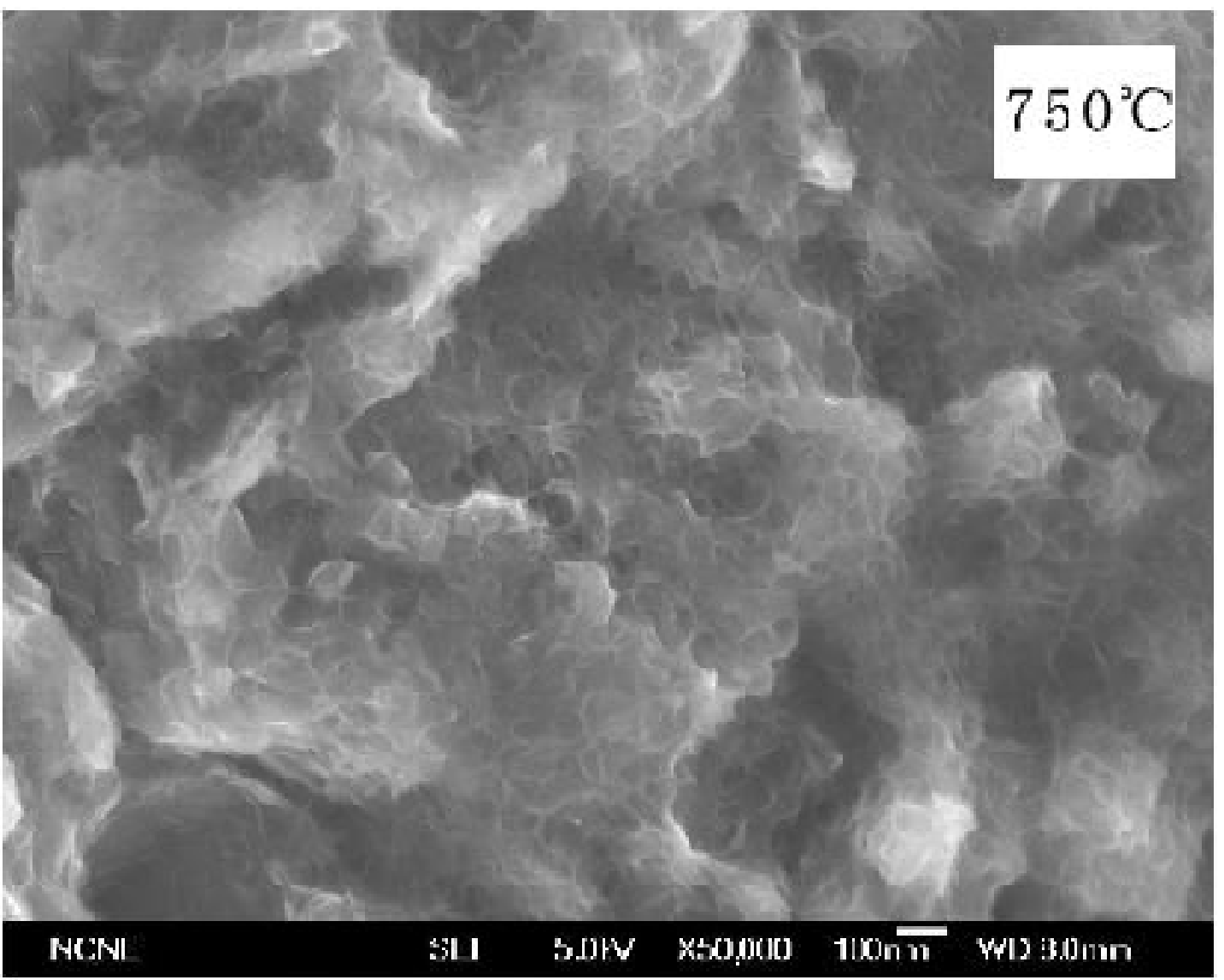}%
\includegraphics[width=200pt]{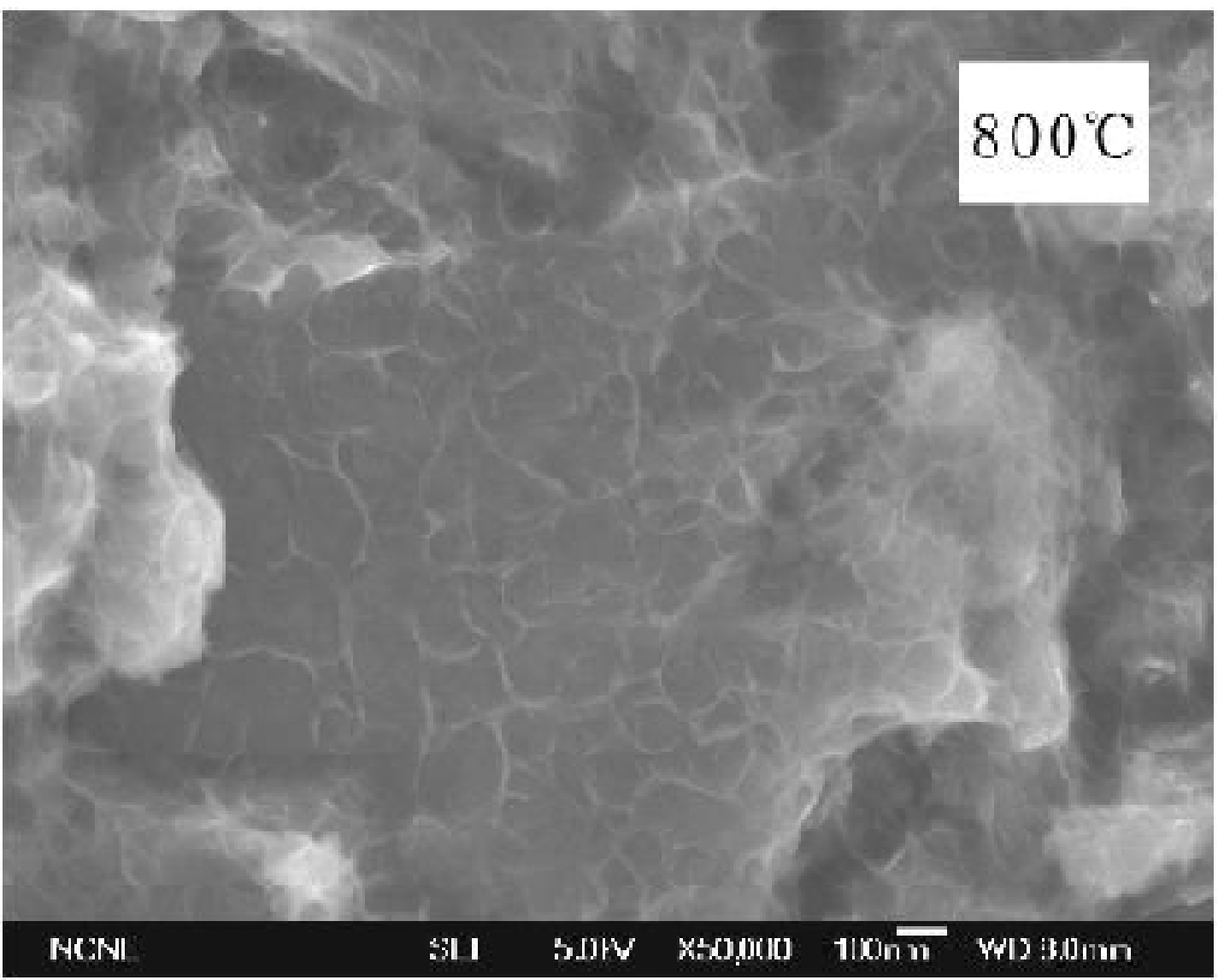}
\end{figure}

\newpage
Fig.3
%\section{EDX for Second phase for 800¡æ sintered sample}
%%%%%%%%%%%%%%%%%%%%%%%%  FIGURE 3 EDX for Second phase for 800¡æ sintered sample  %%%%%%%%%%%%%%%%%%%%%%%%%
\begin{figure}[hp]
\centering
\includegraphics[width=250pt]{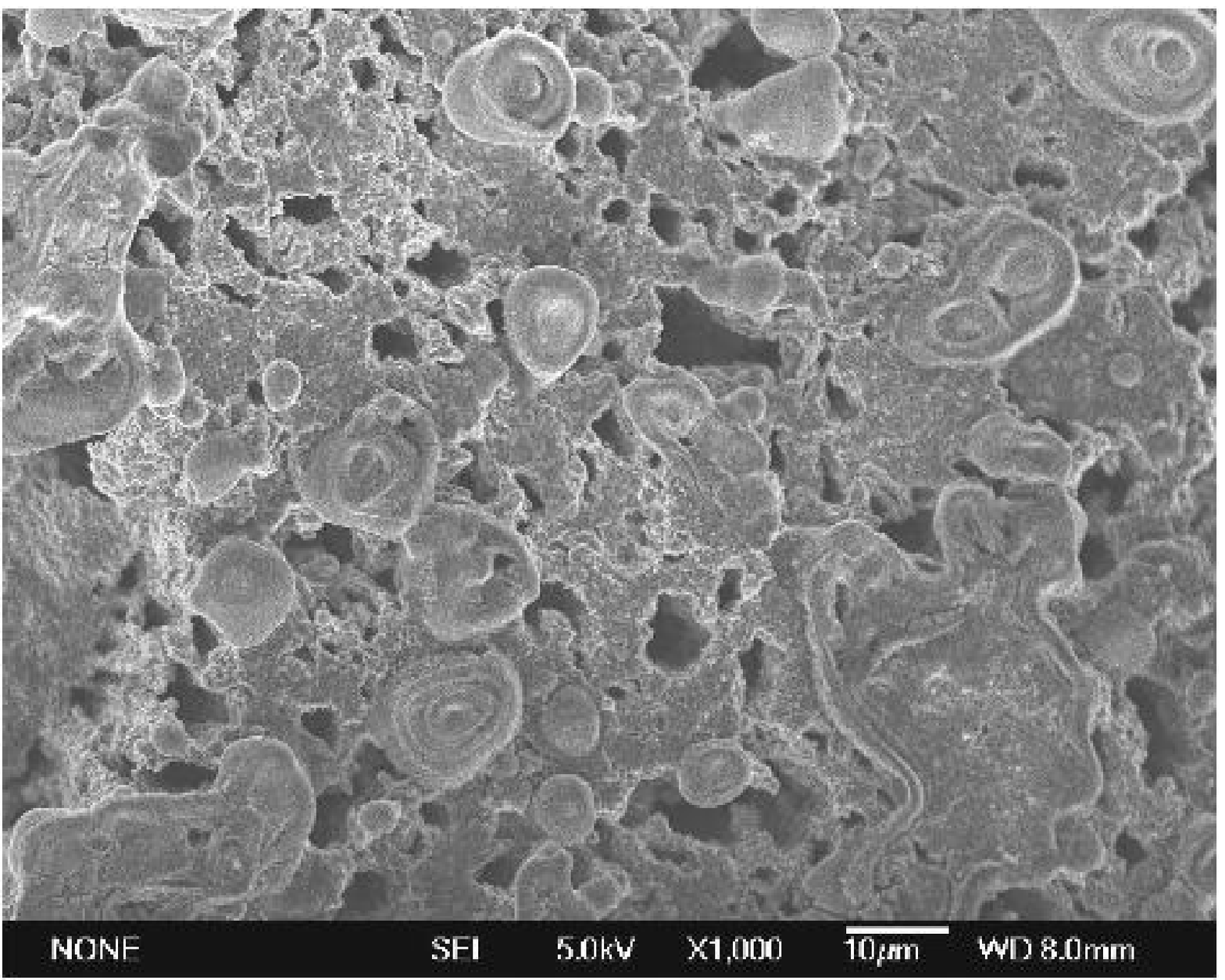}
\end{figure}

\newpage
Fig.4
%\section{Wrap-around distribution of 800¡æ sintered sample}
%%%%%%%%%%%%%%%%%%%%%%%%  FIGURE 4 Wrap-around distribution of 800¡æ sintered sample  %%%%%%%%%%%%%%%%%%%%%%%%%
\begin{figure}[hp]
\centering
\includegraphics[width=200pt]{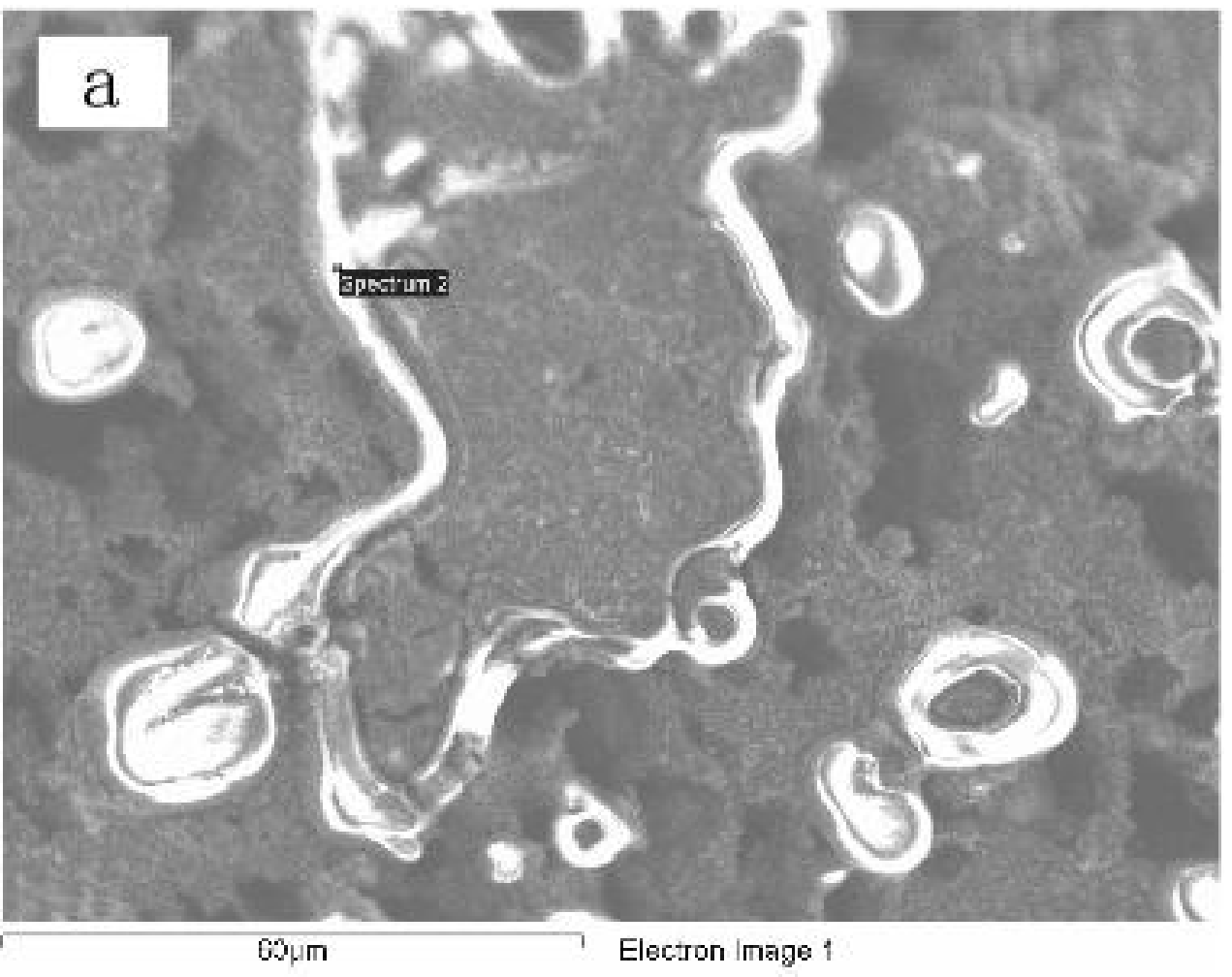}%
\includegraphics[width=200pt]{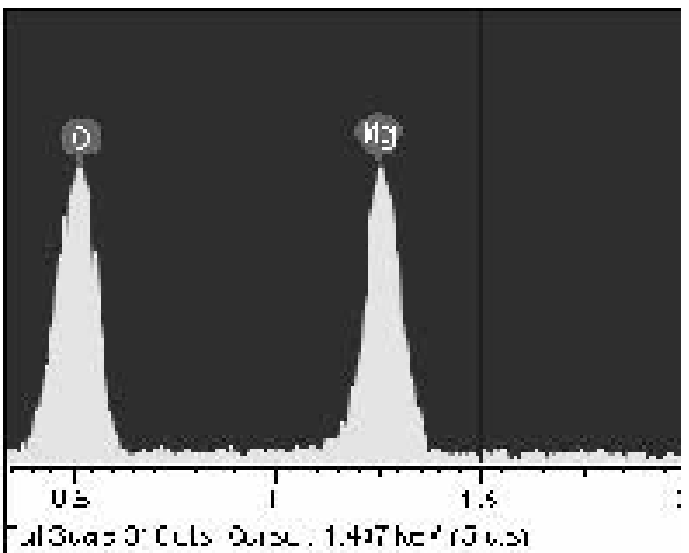}
\includegraphics[width=200pt]{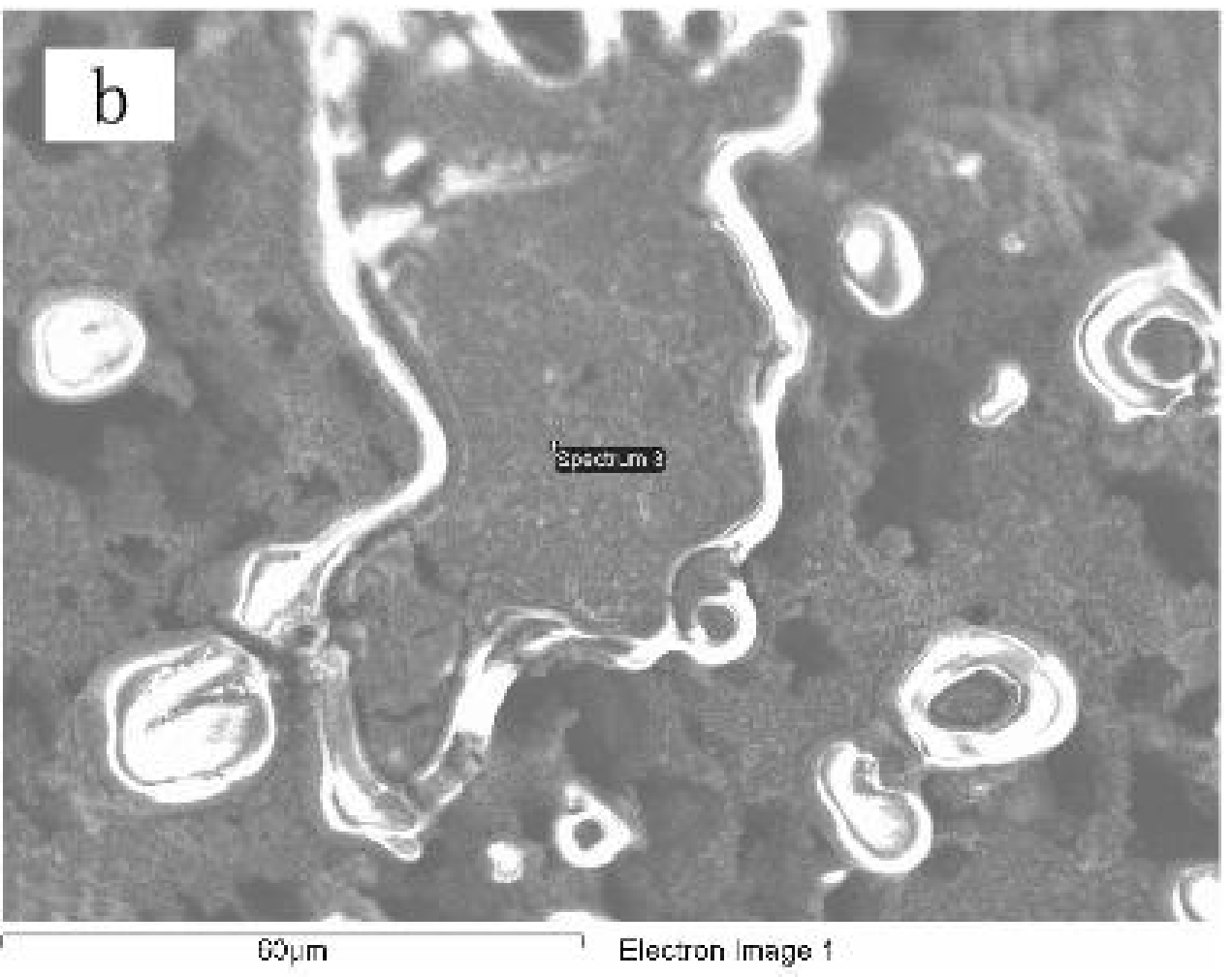}%
\includegraphics[width=200pt]{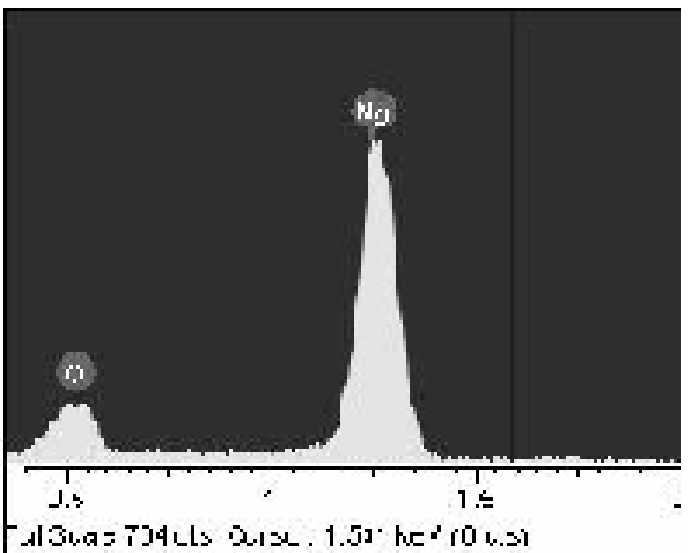}
\end{figure}

\newpage
Fig.5
%\section{R-T}
%%%%%%%%%%%%%%%%%%%%%%%%  FIGURE 5 R-T %%%%%%%%%%%%%%%%%%%%%%%%%
\begin{figure}[hp]
\centering
\includegraphics[width=300pt]{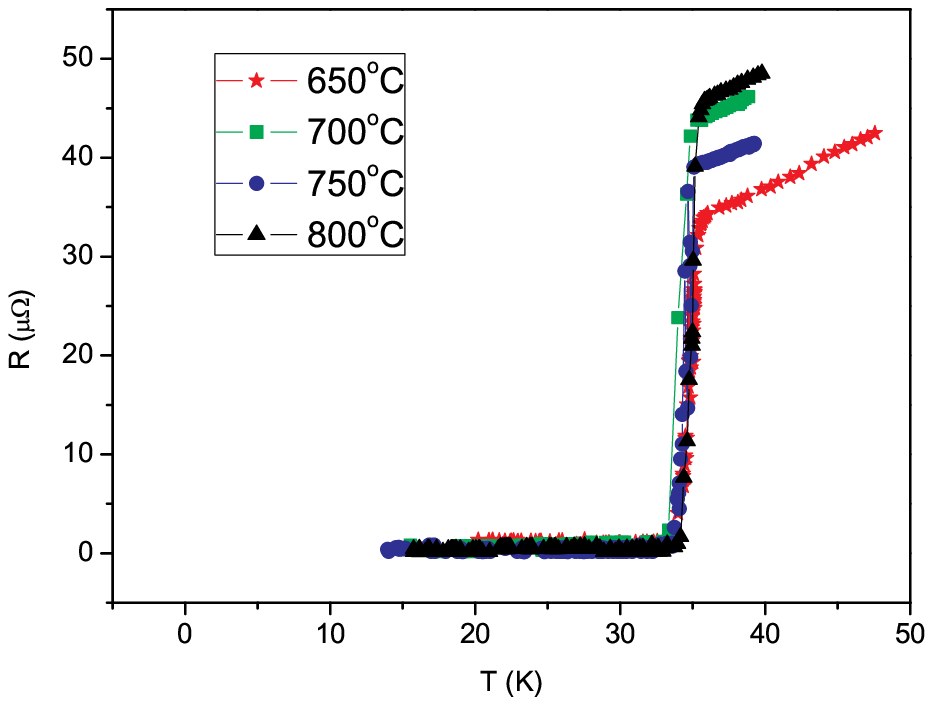}%
\end{figure}

\newpage
Fig.6
%\section{Jc-H}
%%%%%%%%%%%%%%%%%%%%%%%%  FIGURE 6 Jc-H  %%%%%%%%%%%%%%%%%%%%%%%%%
\begin{figure}[hp]
\centering
\includegraphics[width=300pt]{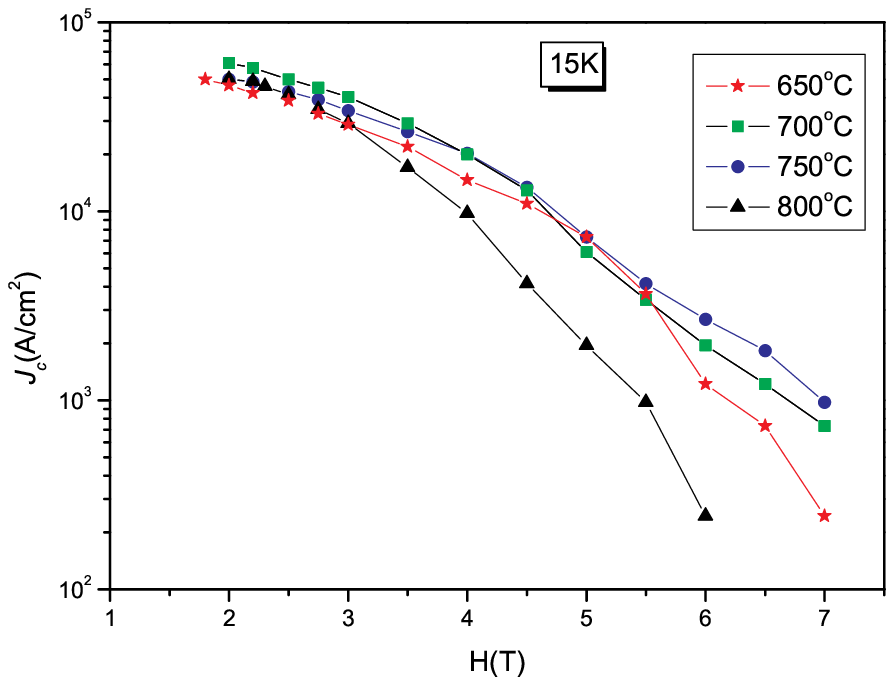}%
\end{figure}


\newpage
\section*{References}
\begin{thebibliography}{10}
\bibitem{book1}V.N. Narozhnyi, G. Fuchs, A, Handstein, A. Gumbel, J. Eckert,
K. Nenkov, D. Hinz, O. Gutfleisch, A. Walte, L. N. Bogacheva, I. E.
Kostyleva, K.-H. Muller, L. Schultz, Int. Conf. on
Superconductivity, CMR \& Related Materials: Novel Trends
(SCRM2002), Giens, France, 1-8 June 2002.

\bibitem{book2}A. Gumbel, O. Perner, J. Eckert, G. Fuchs, K. Nenkov,
K.-H. Muller, and L. Schultz, IEEE TRANSACTIONS ON APPLIED
SUPERCONDUCTIVITY, Vol. 13, No. 2, JUNE 2003.

\bibitem{book3}A. Gumbel, J. Eckert, G. Fuchs, K. Nenkov, K.-H. Muller and L. Schultz,
APPLIED PHYSICS LETTERS, Vol. 80, No. 15, 15 APRIL 2002, pp.2725-2727.

\bibitem{book4}W. Ha¦Âler, C. Roding, C. Fischer, B. Holzapfel, O. Perner, J. Eckert, K. Nenkov
and G. Fuchs, Supercond. Sci. Technol. 16 (2003) pp.281-284.

\bibitem{book5}C. Fischer, C. Roding, W. Ha¦Âler, O. Perner, J. Eckert, K. Nenkov, G. Fuchs,
H. Wendrock, B. Holzapfel and L. Schultz, APPLIED PHYSICS LETTERS,
Vol. 83, No. 9, 1 SEPTEMBER 2003, pp.1803-1805.

\bibitem{book6}O. Perner, J. Eckert, W. Ha¦Âler, C. Fischer, K-H Muller, G. Fuchs, B Holzapfel and
L Schultz, Supercond. Sci. Technol. 17 (2004) pp.1148-1153.

\bibitem{book7}Olaf Perner, Wolfgang Ha¦Âler, Claus Fischer, Gunter Fuchs, Bernhard Holzapfel,
Ludwig Schultz and Jurgen Eckert, IEEE TRANSACTIONS ON APPLIED
SUPERCONDUCTIVITY, VOL. 15, NO. 2, JUNE 2005.

\bibitem{book8}O. Perner, J. Eckert, W. Ha¦Âler, C. Fischer, J. Acker, T. Gemming, G. Fuchs,
B. Holzapfel and L. Schultz, JOURNAL OF APPLIED PHYSICS 97,
056105(2005), pp.1-3.

\bibitem{book9}C. Buzea and T. Yamashita, Supercond. Sci. Technol. 14, R 115 (2001).

\bibitem{book10}V. N. Narozhnyi, G. Fuchs, A. Handstein, A. G¨¹mbel, J. Eckert, K. Nenkov,
D. Hinz, O. Gutfleisch, A. W?lte, L. N. Bogacheva, I. E. Kostyleva,
K.-H. M¨¹ller, and L. Schultz, J. Supercond. 15, 599 (2002)

\end{thebibliography}
\end{document}